%% file: main.tex
\documentclass{article}

\usepackage{arxiv}

\usepackage[utf8]{inputenc} 
\usepackage[T1]{fontenc}    
\usepackage{hyperref}       
\usepackage{url}            
\usepackage{booktabs}       
\usepackage{amsfonts}       
\usepackage{nicefrac}       
\usepackage{microtype}      
\usepackage{lipsum}

\input{solidity-highlighting.tex}

\usepackage{tikz}
\usepackage{amsmath}
\usepackage[normalem]{ulem}

\title{\ProjName{}: A Versatile Framework for Specifying and Verifying Smart Contracts}

\newcommand*\samethanks[1][\value{footnote}]{\footnotemark[#1]}

\author{
  Shaokai Lin
  \thanks{Equal contribution.} \\
  Department of Computer Science\\
  Columbia University\\
  New York, NY 10027 \\
  \texttt{sl4299@columbia.edu} \\
   \And
  Xinyuan Sun\samethanks\\
  CertiK\\
  New York, NY 10018 \\
  \texttt{sxysun@certik.io} \\
    \And
  Jianan Yao \\
  Department of Computer Science\\
  Columbia University\\
  New York, NY 10027 \\
  \texttt{jy3022@columbia.edu} \\
    \And
  Ronghui Gu \\
  Department of Computer Science\\
  Columbia University\\
  New York, NY 10027 \\
  \texttt{ronghui.gu@columbia.edu} \\
}

\input{preamble.tex}
\begin{document}
\maketitle

\begin{abstract}
The growing adoption of smart contracts on
blockchains poses new security risks that can lead to significant monetary loss, while existing approaches
either provide no (or partial) security guarantees for smart contracts
or require huge proof effort.
To address this challenge, we present 
\ProjName{},
a versatile  framework for 
specifying and verifying
industrial-grade smart contracts.
\ProjName{}'s versatile approach  extends previous  efforts with three key contributions: 
(i) an expressive annotation system enabling built-in directives for vulnerability pattern checking, neural-based loop invariant inference, and the verification
of rich properties of real-world smart contracts
(ii) a fine-grained model for the Ethereum Virtual Machine (EVM) that provides low-level execution semantics,
(iii) an IR-level verification framework integrating both SMT solvers and the Coq proof assistant.

We use  \ProjName{} 
to specify and verify security properties
for
12 benchmark contracts and 
a real-world Decentralized Finance (DeFi) smart contract.
Among all 158 specified security properties (in six types),
151 properties can be automatically verified within 2 seconds,
five properties can be automatically verified after moderate modifications,
and two properties are manually proved with around 200 lines of Coq code.
\end{abstract}

\input{sections/1_introduction}

\input{sections/2_overview}

\input{sections/3_annotation}

\input{sections/4_ir}

\input{sections/5_whyml}

\input{sections/6_vc}

\input{sections/7_case_study}

\input{sections/8_related}

\input{sections/9_conclusion}

\bibliographystyle{unsrt}  
\bibliography{main}

\end{document}

%% file: solidity-highlighting.tex
\usepackage{listings, xcolor}

\definecolor{verylightgray}{rgb}{.97,.97,.97}
\definecolor{dark-gray}{gray}{0.2}
\definecolor{mygreen}{HTML}{467E7E}
\definecolor{mygray}{rgb}{0.5,0.5,0.5}
\definecolor{mymauve}{HTML}{AC134D}
\definecolor{myblue}{HTML}{144B7D}
\definecolor{myorange}{HTML}{B25A00}

\lstdefinelanguage{Solidity}{
	keywords=[1]{anonymous, assembly, assert, balance, break, call, callcode, case, catch, class, constant, continue, constructor, contract, debugger, default, delegatecall, delete, do, else, emit, event, experimental, export, external, false, finally, for, function, gas, if, implements, import, in, indexed, instanceof, interface, internal, is, length, library, log0, log1, log2, log3, log4, memory, modifier, new, payable, pragma, private, protected, public, pure, push, require, return, returns, revert, selfdestruct, send, solidity, storage, struct, suicide, super, switch, then, this, throw, transfer, true, try, typeof, using, view, while, with, addmod, ecrecover, keccak256, mulmod, ripemd160, sha256, sha3}, 
	keywordstyle=[1]\color{myblue!70}\bfseries,
	keywords=[2]{address, bool, byte, bytes, bytes1, bytes2, bytes3, bytes4, bytes5, bytes6, bytes7, bytes8, bytes9, bytes10, bytes11, bytes12, bytes13, bytes14, bytes15, bytes16, bytes17, bytes18, bytes19, bytes20, bytes21, bytes22, bytes23, bytes24, bytes25, bytes26, bytes27, bytes28, bytes29, bytes30, bytes31, bytes32, enum, int, int8, int16, int24, int32, int40, int48, int56, int64, int72, int80, int88, int96, int104, int112, int120, int128, int136, int144, int152, int160, int168, int176, int184, int192, int200, int208, int216, int224, int232, int240, int248, int256, mapping, string, uint, uint8, uint16, uint24, uint32, uint40, uint48, uint56, uint64, uint72, uint80, uint88, uint96, uint104, uint112, uint120, uint128, uint136, uint144, uint152, uint160, uint168, uint176, uint184, uint192, uint200, uint208, uint216, uint224, uint232, uint240, uint248, uint256, var, void, ether, finney, szabo, wei, days, hours, minutes, seconds, weeks, years},	
	keywordstyle=[2]\color{myorange!70}\bfseries,
	keywords=[3]{block, blockhash, coinbase, difficulty, gaslimit, number, timestamp, msg, data, gas, sender, sig, value, now, tx, gasprice, origin},	
	keywordstyle=[3]\color{violet}\bfseries,
	identifierstyle=\color{black},
	sensitive=false,
	comment=[l]{//},
	morecomment=[s]{/*}{*/},
	commentstyle=\color{mygray}\ttfamily,
	stringstyle=\color{mymauve}\ttfamily,
	morestring=[b]',
	morestring=[b]"
}

\lstdefinestyle{solidity}{
  language=Solidity,
  backgroundcolor=\color{white},
  basicstyle=\linespread{0.9}\ttfamily\footnotesize,
  breakatwhitespace=false,
  breaklines=true,                 
  numberstyle=\tiny\color{gray}, 
  keepspaces=true, 
  showspaces=false,                
  showstringspaces=false,          
  showtabs=false,                  
  tabsize=1,  
  emph = {forall},
  emphstyle=\bfseries\color{mymauve},%
}

%% file: preamble.tex
\usepackage{amssymb}
\usepackage{amsmath}

\usepackage{booktabs}

\usepackage{scrextend}

\usepackage{ifpdf}
\usepackage{graphicx}

\usepackage{color} 

\usepackage{mathtools}

\usepackage{epstopdf}

\usepackage[toc,page]{appendix}

\usepackage{listings}
\usepackage{why3lang}

\usepackage{svg}

\definecolor{verylightgray}{rgb}{.97,.97,.97}
\definecolor{dark-gray}{gray}{0.2}
\definecolor{mygreen}{HTML}{467E7E}
\definecolor{mygray}{rgb}{0.5,0.5,0.5}
\definecolor{mymauve}{HTML}{AC134D}
\definecolor{myblue}{HTML}{144B7D}
\definecolor{myorange}{HTML}{B25A00}

\lstdefinelanguage{Coq}{ 
mathescape=true,
texcl=false, 
morekeywords=[1]{Section, Module, End, Require, Import, Export,
  Variable, Variables, Parameter, Parameters, Axiom, Hypothesis,
  Hypotheses, Notation, Local, Tactic, Reserved, Scope, Open, Close,
  Bind, Delimit, Definition, Let, Ltac, Fixpoint, CoFixpoint, Add,
  Morphism, Relation, Implicit, Arguments, Unset, Contextual,
  Strict, Prenex, Implicits, Inductive, CoInductive, Record,
  Structure, Canonical, Coercion, Context, Class, Global, Instance,
  Program, Infix, Theorem, Lemma, Corollary, Proposition, Fact,
  Remark, Example, Proof, Goal, Save, Qed, Defined, Hint, Resolve,
  Rewrite, View, Search, Show, Print, Printing, All, Eval, Check,
  Projections, inside, outside, Def},
morekeywords=[2]{forall, exists, exists2, fun, fix, cofix, struct,
  match, with, end, as, in, return, let, if, is, then, else, for, of,
  nosimpl, when},
morekeywords=[3]{Type, Prop, Set, true, false, option},
morekeywords=[4]{pose, set, move, case, elim, apply, clear, hnf,
  intro, intros, generalize, rename, pattern, after, destruct,
  induction, using, refine, inversion, injection, rewrite, congr,
  unlock, compute, ring, field, fourier, replace, fold, unfold,
  change, cutrewrite, simpl, have, suff, wlog, suffices, without,
  loss, nat_norm, assert, cut, trivial, revert, bool_congr, nat_congr,
  symmetry, transitivity, auto, split, left, right, autorewrite},
morekeywords=[5]{by, done, exact, reflexivity, tauto, romega, omega,
  assumption, solve, contradiction, discriminate},
morekeywords=[6]{do, last, first, try, idtac, repeat},
morecomment=[s]{(*}{*)},
showstringspaces=false,
morestring=[b]",
morestring=[d],
tabsize=3,
extendedchars=false,
sensitive=true,
breaklines=false,
basicstyle=\small,
captionpos=b,
columns=[l]flexible,
identifierstyle={\ttfamily\color{black}},
keywordstyle=[1]{\ttfamily\color{violet}},
keywordstyle=[2]{\ttfamily\color{mygreen}},
keywordstyle=[3]{\ttfamily\color{myblue}},
keywordstyle=[4]{\ttfamily\color{myblue}},
keywordstyle=[5]{\ttfamily\color{red}},
stringstyle=\ttfamily,
commentstyle={\ttfamily\color{mygreen}},
literate=
    {\\forall}{{\color{dkgreen}{$\forall\;$}}}1
    {\\exists}{{$\exists\;$}}1
    {<-}{{$\leftarrow\;$}}1
    {=>}{{$\Rightarrow\;$}}1
    {==}{{\code{==}\;}}1
    {==>}{{\code{==>}\;}}1
    {->}{{$\rightarrow\;$}}1
    {<->}{{$\leftrightarrow\;$}}1
    {<==}{{$\leq\;$}}1
    {\#}{{$^\star$}}1 
    {\\o}{{$\circ\;$}}1 
    {\@}{{$\cdot$}}1 
    {\/\\}{{$\wedge\;$}}1
    {\\\/}{{$\vee\;$}}1
    {++}{{\code{++}}}1
    {~}{{\ }}1
    {\@\@}{{$@$}}1
    {\\mapsto}{{$\mapsto\;$}}1
    {\\hline}{{\rule{\linewidth}{0.5pt}}}1
}[keywords,comments,strings]

\lstnewenvironment{coq}{\lstset{language=Coq}}{}

\usepackage{subfigure}

\usepackage{scalerel}

\usepackage{graphicx}

\usepackage{tikz}
\usetikzlibrary{shapes,arrows,automata,positioning,cd}
\tikzset{
  cfgedge/.style   = {black, ->, >=stealth},
  forward/.style = { blue, ->, >=angle 45},
  backward/.style = { red, densely dashed, ->, >=latex' },
  backwardleft/.style = { red, densely dashed, <-, >=latex' },
}

\usepackage{xcolor, colortbl}
\newcommand{\red}{\cellcolor[HTML]{F88379}}
\newcommand{\green}{\cellcolor[HTML]{99D18F}}

\usetikzlibrary{fit,calc}

\colorlet{pink}{red!40}
\colorlet{cyan}{cyan!60}

\usepackage{prettyref}

\newrefformat{thm}{Theorem~\ref{#1}}
\newrefformat{lem}{Lemma~\ref{#1}}
\newrefformat{cha}{Chapter~\ref{#1}}
\newrefformat{sec}{Section~\ref{#1}}
\newrefformat{tab}{Table~\ref{#1}}
\newrefformat{fig}{Figure~\ref{#1}}
\newrefformat{alg}{Algorithm~\ref{#1}}
\newrefformat{exa}{Example~\ref{#1}}
\newrefformat{def}{Definition~\ref{#1}}
\newrefformat{li}{Line~\ref{#1}}

\usepackage{mathpartir}

\usepackage[ruled,vlined,linesnumbered]{algorithm2e}

\definecolor{light-gray}{gray}{0.9}
\definecolor{light-pink}{rgb}{0.858, 0.188, 0.478}

\newcommand{\Omit}[1]{}

\usepackage[flushleft]{threeparttable} 
\usepackage{booktabs}
\usepackage[labelfont={bf}]{caption}
\newcommand{\repthanks}[1]{\textsuperscript{\ref{#1}}}
\makeatletter
\patchcmd{\maketitle}
  {\def\thanks}
  {\let\repthanks\repthanksunskip\def\thanks}
  {}{}
\patchcmd{\@maketitle}
  {\def\thanks}
  {\let\repthanks\@gobble\def\thanks}
  {}{}
\newcommand\repthanksunskip[1]{\unskip{}}
\makeatother
\newcommand{\ProjName}{SciviK}

\makeatletter
\renewcommand{\paragraph}{%
  \@startsection{paragraph}{4}%
  {\z@}{1ex \@plus 1ex \@minus .2ex}{-1em}%
  {\normalfont\normalsize\bfseries}%
}
\makeatother

\definecolor{fixmeColor}{RGB}{0,145,225}

\newtheorem{theorem}{Theorem}

\newcommand{\para}[1]{\vspace{5pt}\noindent{}\textbf{#1}\ }

%% file: sections/1_introduction.tex
\section{Introduction}
\label{sec:Introduction}

Blockchain and other distributed ledger technologies enable consensus over global computation to be applied to situations where decentralization and security are critical~\cite{zheng2018blockchain}. 
Smart contracts are decentralized programs on the blockchain  
that encode the logic of transactions and businesses, enabling a new form of collaboration --- rather than requiring a trusted third party,
users only need to 
trust that the smart contracts faithfully encode the transaction logic~\cite{Ethereum_2020}.
However, the smart contract implementations are
not trustworthy due to program errors and, by design, are difficult to change once deployed~\cite{nakamoto2009, zheng2018blockchain}, posing new security risks.
Million dollars' worth of digital assets have been stolen every week
due to security vulnerabilities in smart contracts~\cite{atzei2017survey}.

Techniques such as
static analysis and formal verification
have been applied to 
improve smart contracts' security
and ensure that given contracts
satisfy desired properties~\cite{tsankov2018securify,luu2016making,nehai2018model,sergey2018temporal,feist2019slither}.
While promising, existing  efforts still suffer
fundamental limitations.

There is a long line of work to improve
 smart contracts' security
using highly automated techniques,
such as security pattern matching \cite{tsankov2018securify}, symbolic execution \cite{luu2016making}, and model checking \cite{nehai2018model, sergey2018temporal}. These efforts can only deal with pre-defined vulnerability patterns
and do not support contract-specific properties.
For example, a smart contract implementing 
a decentralized exchange may need to
ensure that traders are always provided
with the optimal price.
Such a property is related to the underlying financial model and
cannot be verified using existing automated approaches.
Furthermore, even for pre-defined patterns,
these highly automated techniques
may still fail to provide any security guarantee. 
For example, the accuracy rates of Securify~\cite{tsankov2018securify} and Oyente~\cite{luu2016making} are  62.05\% and 54.68\%, respectively, on various benchmarks~\cite{parizi2018empirical},
making them ineffective and impractical 
in securing complex smart contracts.

Previous efforts focusing on developing
mechanized proofs for  smart contracts
can provide security guarantees
but
often require substantial manual effort
to write boilerplate specifications, 
infer invariants for every loop or recursive procedure,
and implement proofs with limited automation 
support~\cite{hildenbrandt2018kevm,hajdu2020smt,annenkov2019towards},
limiting their application to industry-grade smart contracts.

Most of the existing techniques
on securing smart contracts,
regardless of their approach,
work
either at the source code level or 
at the bytecode level.
However, each of these two target levels
has its own drawbacks.
On the one hand, 
techniques working at the source code level
are hard to adapt to 
the rapid development 
of smart contract's toolchain.
Programming languages for writing smart contracts
are still under active and iterative development
and usually do not have a formal (or even informal) semantics definition.
Solidity, the official language for Ethereum smart contracts,
has released 84 versions from Aug 2015 to Jan 2021 and still does not have a stable formal semantics. 
On the other hand, due to a lack of program structure and the exposure of low-level EVM details, techniques working at the bytecode level usually only support a limited set of security properties while not allowing the writing of manual proofs for the properties that cannot be handled by automated techniques, such as the ones related to financial models~\cite{luu2016making, mueller2017mythril, hildenbrandt2018kevm, feist2019slither}.

To address the above challenges, we present \ProjName{}, a versatile  framework 
enabling the formal verification
of smart contracts at an intermediate representation (IR) level
with respect to a rich set of pre-defined and user-defined
properties expressed by lightweight source-level annotations.

\ProjName{}  introduces a source-level annotation system, with which users can write expressive specifications by directly annotating the smart contract source code.
The annotation system  supports a rich set of built-in directives, 
including vulnerability pattern checking (\texttt{@check}), neural-based loop invariant inference (\texttt{@learn}),
and exporting proof obligations to the Coq proof assistant (\texttt{@coq}). 
The annotation system offers the flexibility between using automation techniques to reduce the proof burden and using an interactive proof assistant to develop complex manual proofs. 

\ProjName{} parses and compiles annotated source programs
into annotated programs in Yul~\cite{yul_docs}, a stable intermediate representation for all programming languages used by the Ethereum ecosystem.
Such a design choice unifies the verification of smart contracts
written in various languages and various versions.

We develop an IR-level verification engine on top of WhyML~\cite{bobot2011why3}, 
an intermediate verification language,
to encode the semantics of annotated Yul IR programs.
To make sure that the verified
guarantees hold on EVM,
\ProjName{} formalizes a high-granularity EVM execution model 
in WhyML,
with which verification conditions (VCs)  that respect EVM behaviors
can be generated from annotated IR programs
and discharged to
a suite of SMT solvers (Z3~\cite{TACAS:deMB08}, Alt-Ergo~\cite{bobot2013alt}, CVC4~\cite{barrett2011cvc4}, etc.) and the Coq proof assistant~\cite{barras1997coq}.

To evaluate \ProjName{},
we study the smart contracts of
167 real-world projects
audited by CertiK~\cite{ctk2020}
and characterize common security properties
into six types.
We then use  \ProjName{} 
to specify and verify
12 benchmark contracts and 
a real-world Decentralized Finance (DeFi) smart contract
with respect to these six types of security properties.
Among all 158 security properties (in six types)
specified for the evaluated smart contracts,
151 properties can be automatically verified within 2 seconds,
five properties can be automatically verified after moderate retrofitting
to the generated WhyML programs,
and two properties are manually proved with around 200 lines of Coq code.

In summary, this paper 
makes the following technical contributions:
\begin{itemize}
  \item An expressive and lightweight annotation system for specifying built-in
  and user-defined properties of smart contracts;
  \item A high-granularity EVM model implemented in WhyML;
  \item An IR-level verification framework combining both SMT solvers
  and the Coq proof assistant to maximize the verification ability and flexibility;
  \item Evaluations showing that \ProjName{} is effective and practical
  to specify and verify all six types of security properties
  for real-world smart contracts.
\end{itemize}

%% file: sections/2_overview.tex
\section{Overview}
\label{sec:Overview}

To give an overview of \ProjName{}, we use a simplified version of the staked voting smart contract, which is commonly used in DeFi,
as a running example (see Figure~\ref{code:votingsol} for the Solidity contract).
This contract allows participants on the blockchain to collectively vote for proposals by staking their funds. 
Contracts like this, such as the Band Protocol~\cite{srinawakoon2019} staked voting application, have managed more than \$150 million worth of cryptocurrencies. However, many bugs such as Frozen Funds~\cite{atzei2017survey} were frequently discovered in such voting-related multi-party contracts due to their complex business logic. Here, we intentionally insert three common vulnerabilities to this voting contract, respectively, on lines~12-13, 15, and 31, which cover vulnerabilities including reentrancy attacks~\cite{reentrancy_solidity_soln}, calculation mistakes, and incorrect data structure operations. 

\begin{figure}[t]
\begin{center}
\begin{lstlisting}[style=solidity]
/* @meta forall i: address, stakers[i] != 0x0 */
contract StakedVoting {
 // ... variable declarations and functions ...
 
 /* @post userVoted[msg.sender] -> revert
  * @post ! old userVoted[msg.sender] ->
  *  stakes[msg.sender] = old stakes[msg.sender] + stake
  * @check reentrancy */
 function vote(uint256 choice, uint256 stake, address token) public {
  if (userVoted[msg.sender]) { revert(); }
  rewards[msg.sender] += _lotteryReward(random(100));
  bool transferSuccess = ERC777(token).
                         transferFrom(msg.sender,address(this), stake);
  /* @assume transferSuccess */
  stakes[msg.sender] *= stake; 
  // ... the voting procedure and adding stakes to the stake pool ...
  userVoted[msg.sender] = true; past_stakes.push(stake);
  rebalanceStakers(); // sort the updated list of voters
 }

 /* @coq @pre sorted_doubly_linked_list stakers
  * @coq @post sorted_doubly_linked_list stakers */
 function rebalanceStakers() internal {
  if (stakers[msg.sender] == address(0)) {
   /* @assert (prevStakers[msg.sender] == address(0)); */
   address prevStaker = HEAD;
   for (address x; x != END; prevStaker = x) {
    // ... add msg.sender node to right position in stakers
   }
  } else {
   // delete old msg.sender node in stakers
   delete stakers[msg.sender];
   delete prevStakers[msg.sender];
   rebalanceStakers();
  }
 }
 
 function _lotteryReward(uint256 n) pure internal returns (uint) {
  uint x = 10; uint y = 0;
  // The user can also use "@inv" to manually provide a loop invariant.
  /* @learn x y
   * @pre n < 100
   * @post y < 10
   * @post x >= n */
  while (x < n) { x += 1; y += x ** 2; }
  return y;
 }
}
\end{lstlisting}
\end{center}
\vspace{-10pt}
\caption{A simplified staked voting contract in Solidity with three vulnerabilities.}
\vspace{-5pt}
\label{code:votingsol}
\end{figure}

\begin{figure*}[t]
  \centering
  \includegraphics[width=0.6\textwidth]{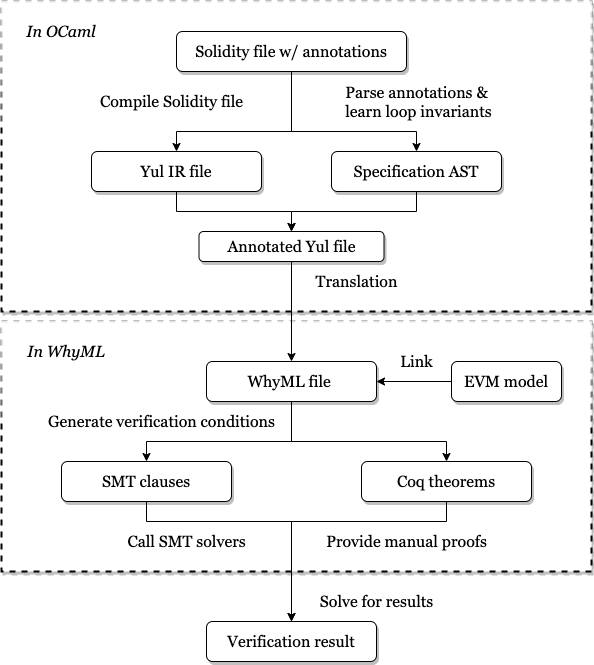}
  \caption{Architecture of \ProjName{} verification framework.}
  \label{fig:arch}
\end{figure*}

To demonstrate the workflow of \ProjName{} and its architecture (see Figure~\ref{fig:arch}), we 
verify the correctness of the
toy voting contract shown in Figure~\ref{code:votingsol} 
by finding the three inserted vulnerabilities and proving that the following theorems hold: 

\begin{itemize}
    \item $\mathcal{T}_1$ (\textit{On-Chain Data Correctness}): The data structure storing voted results will always reflect actual voter operations.
    \item $\mathcal{T}_2$ (\textit{Access Control}): Only authenticated users that have neither voted nor collected lottery rewards before can vote. 
    \item $\mathcal{T}_3$ (\textit{Correctness of Voting Procedure}): The voting procedure itself is correct, i.e., stake accounts and user decisions are correctly updated after each vote.
\end{itemize}

\para{Step 1: Writing annotations.}
\ProjName{} requires users to provide the desired properties in the comments using its annotation system, which supports arithmetic operators, comparison operators, first-order logic operators, and built-in directives, such that a rich set of properties can be easily defined.
For example, $\mathcal{T}_2$, which states that people who have voted cannot vote again, can be formally specified by the \texttt{@post} tag on line~5 that if the user does not meet the requirement, the function reverts. 
This is an example of a function-level annotation that only holds for the annotated function call.
Further, the \texttt{@meta} specification on line~1 asserts a contract-level invariant that the array of stakers will never contain the address \texttt{0x0} (an invalid address for \texttt{msg.sender}) at any observable point in the contract execution.
The \texttt{@post} condition on line~6-7 ensures that if a qualified voter votes successfully, the voter's stakes increase by the number of staked tokens after the vote.
The \texttt{@check} directive on line~8 instantiates a vulnerability pattern check that checks for any reentrancy attacks in the function.
These specifications together contribute to the validity of $\mathcal{T}_3$.

The \ProjName{} annotation system provides users the flexibility to reason about different properties at different abstraction levels.
For example, one can choose to abstract the hashmap operators in Yul to functional data structures and inductive predicates (see Section~\ref{sec:generatingir} for details). 
Another example is the specifications annotated with \texttt{@coq} prefix on line~21-22, where it aims to prove $\mathcal{T}_1$ by asserting that, if \texttt{stakers} is a sorted doubly linked list, it remains a sorted doubly linked list
after the insertion of a new voter node. This annotation enables \ProjName{}
to port the verification condition of \texttt{rebalanceStakers} to Coq.

\para{Step 2: Parsing annotated programs into annotated Yul IR.}
\ProjName{} parses the program with user-provided annotations and then uses the Continuous Logic Networks (CLN)~\cite{ryan2019cln2invold,ryan2019cln2inv} to automatically generate numeric loop invariants based on \texttt{@learn} directive. For example, the annotation on line 41 indicates that we expect \ProjName{} to automatically infer a loop invariant for our pseudo-random \texttt{\_lotteryReward} function.

After parsing annotated Solidity, \ProjName{} compiles the smart contract using the \texttt{solc} compiler and outputs annotated Yul IR
of the contract with embedded EVM opcodes. 
Based on the generated Yul IR, \ProjName{} compiles annotations and learned invariants to IR-level annotations and insert them into the generated Yul IR.
Since Yul is intermixed with low-level EVM opcodes and manipulates non-local variables via pointer arithmetic and hashing, at this step, \ProjName{} also needs to map all non-local variables
to corresponding memory and storage operations for later data type abstraction.
For example, 
the Yul IR snippet that contains the IR-level annotation
corresponding to the \texttt{@post} tag on line~6-7 is shown below:

\begin{lstlisting}[language = Why3, mathescape=true]
/* ...
 * @post 
 *   !old read_from_storage_split_offset_0_t_bool(
 *          mapping_index_access_t_mapping_t_address_t_bool_of_t_address(
 *            0x05, caller())) 
 *   -> read_from_storage_split_offset_0_t_uint256(
 *        mapping_index_access_t_mapping_t_address_t_uint256_of_t_address(
 *          0x06, caller())) 
 *      = old read_from_storage_split_offset_0_t_uint256(
 *              mapping_index_access_t_mapping_t_address_t_uint256_of_t_address(
 *                0x06, caller())) 
 *        + vloc_stake_226
 * ... */
function fun_vote_277(vloc_choice_224, vloc_stake_226, vloc_token_228) {
   ... the voting procedure ...
}
\end{lstlisting}

\para{\bf Step 3: Translating annotated Yul IR to WhyML.} 
In this step, \ProjName{} translates the compiled Yul IR into WhyML~\cite{bobot2011why3} with specifications inserted at designated locations. 
During translation, each Yul IR function, with the IR-level annotations, is mapped to a semantically equivalent definition in WhyML. The patterns specified by \texttt{@check} are also expanded to concrete specifications in WhyML. \ProjName{} further links the generated WhyML to an EVM model that contains the semantic definitions of the EVM opcodes implemented in WhyML.
A WhyML snippet of the translated example contract is shown below:
\vspace{5pt}
\begin{lstlisting}[language = Why3,  mathescape=true]
let ghost function fun_vote_277 (st_c: evmState)  (vloc_choice_224: int)  
 (vloc_stake_226: int) (vloc_token_228: int) : evmState
 ...
 ensures { !(Map.get st_c.map_0x05 st_c.msg.sender) 
           -> (Map.get result.map_0x06 result.msg.sender)     
              = (Map.get st_c.map_0x06 st_c.msg.sender) + vloc_stake_226 }
 = let _r: ref int = ref 0 in
   let ghost st_g: ref evmState = ref st_c in
   try
     begin
       ... the voting procedure ...
     end;
     st_g := (setRet !st_g !_r);
     raise Ret
   with Ret -> (!st_g) 
   end
\end{lstlisting}
 
\para{Step 4: Generating VCs and proofs.}
\ProjName{}  generates the VCs
using the WhyML programs and the EVM model,
and then discharges the VCs to
SMT solvers.
In case SMT solvers fail to solve the VCs automatically, 
the user can choose to port these proof obligations to the Coq
proof assistant with \ProjName{}  and prove them manually
in Coq's interactive proof mode.

We now dive into the three vulnerabilities in the example above. The first one is the reentrancy bug on line~12-13 that can be detected by \ProjName{}'s \texttt{@check reentrancy} directive on line 8. The reentrancy attack could happen as follows. On line 17, the state variable \texttt{userVoted} is updated after an invocation to an external function, namely, the user-provided ERC777~\cite{erc777} token contract's function \texttt{transferFrom}. This function's semantics is to transfer \texttt{stake} amount of token from the voter to the contract. However, since the ERC777 token contract's address is provided by the user, the implementation of its \texttt{transferFrom} function could be malicious. For example, \texttt{transferFrom} could call the \texttt{vote} function again and claim reward multiple times on line~12-13, bypassing the check of \texttt{userVoted} at line 10. This kind of exploits on ERC777-related reentrancy has caused tens of millions of capital loss on DeFi protocol Lendf.me~\cite{lendf}. The exploit can be fixed if we place the call to the ERC777 token immediately after line 18.
Furthermore, \ProjName{}'s discharge of the VC generated by annotation on line~6-7 does not pass and the SMT solver gives a counter-example, where we find that the stakes of a voter should be added rather than multiplied.
During the process of manually proving the annotations on line 21 and 22, we find that Coq always fails to reason the case where an old voter node already existed in the doubly linked list \texttt{stakers}. After inspecting the implementation of \texttt{rebalanceStakers}, we find that there should have been a line of \texttt{stakers[prevStakers[msg.sender]] = stakers[msg.sender];} immediately after line 31 to connect the original voter node's predecessor and successor before their deletions. 

\para{Limitations and assumptions.}
\ProjName{}  does not support 
the generation of loop invariants and intra-procedural assertions
for Coq.
Among 158 security properties verified for evaluated smart contracts,
five properties require manual retrofitting for the generated WhyML programs.
This retrofitting step can be automated and is left for future work.
\ProjName{} trusts the correctness of the  
(1) compiler backend that compiles
Yul IR programs to EVM bytecode, 
(2) the translation from source-level annotations
to IR-level annotations,
(3) the translation from annotated Yul IR programs to annotated WhyML programs, 
(4) the generation of verification conditions from WhyML programs,
(5) SMT solvers,
and (6) the Coq proof checker. 
We do not need to trust
the compiler front-end that parses
the source programs to IR programs.

%% file: sections/3_annotation.tex
\section{The Annotation System} 
\label{sec:annotationsystem}

\begin{figure}[t]
\begin{center}
$
{\small
\begin{array}{llllllllllll}
(\textit{IdPrefix}) &\textit{idp}  \ ::= \ 
\texttt{old} \ | \ \epsilon \\
(\textit{Ident}) &\textit{idnt}  \ ::= \ 
\texttt{x} \ | \ \texttt{result} \ | \ \textit{idp} \ \texttt{x} \\
(\textit{Quant}) &\textit{qunt}  \ ::= \ 
\texttt{forall} \ | \ \texttt{exists} \\
(\textit{Pattern}) &\textit{pat} \ \ \ ::= \ 
\texttt{overflow} \  | \
\texttt{re-entrancy} \\ 
& \ \ \ \ \ \ \ \ \ \ \ \ \  | \ \texttt{timestamp} \\
(\textit{Status}) &\textit{stat} \ ::= \
\texttt{return} \ | \ \texttt{revert} \\
(\textit{Prefix}) &\textit{p} \ \ \ \ \ ::= \
\texttt{@coq} \ | \ \texttt{old} \ | \ \epsilon \\
(\textit{Form}) & \textit{form} \ ::= \ \ \texttt{e} \ | \ \textit{stat} \ | \ \texttt{!} \ \textit{form} \ | \ \textit{form} \ \Rightarrow \ \textit{form} \\
& \ \ \ \ \ \ \ \ \ \ \ \ \ \ \ | \
\textit{form} \ \wedge \textit{form} \ | \ \textit{form} \ \vee \ \textit{form} \\
& \ \ \ \ \ \ \ \ \ \ \ \ \ \ \ | \
\textit{qunt} \ \overline{\textit{idnt : t}} \ , \ \textit{form}  \\
(\textit{Spec}) &\textit{spec} \ ::= \
p \ \texttt{@pre} \ \{ \ \textit{form} \ \} \ | \ 
p \ \texttt{@post} \ \{ \ \textit{form} \ \} \\
& \ \ \ \ \ \ \ \ \ \ \ \ | \
p \ \texttt{@meta} \ \{ \ \textit{form} \ \} \ | \
\texttt{@inv} \ \{ \ \textit{form} \ \} \\
& \ \ \ \ \ \ \ \ \ \ \ \ | \
\texttt{@assume} \ \{ \ \textit{form} \ \} \ | \
\texttt{@assert} \ \{ \ \textit{form} \ \} \ | \\
& \ \ \ \ \ \ \ \ \ \ \ \ | \
\texttt{@check} \ \textit{pat} \\
& \ \ \ \ \ \ \ \ \ \ \ \ | \
\texttt{@learn} \ \overline{\texttt{x}} 

\end{array}
}
$
\end{center}
\vspace{-5pt}
\caption{The syntax of \ProjName{}'s annotation system.}
\label{fig:mach:syntaxsck}
\end{figure}

As shown in Figure~\ref{fig:mach:syntaxsck}, the annotation system of \ProjName{} consists of a large set of directives for constructing specifications and verification conditions. 
\ProjName{} supports different types of annotations, including pre-condition (\texttt{@pre}), post-condition (\texttt{@post}), loop invariant (\texttt{@inv}), global invariant that holds true at all observable states (\texttt{@meta}), assumption (\texttt{@assume}), and assertion (\texttt{@assert}). We illustrate the use of the above directives in the overview example (see Figure~\ref{code:votingsol}).

In Figure~\ref{fig:mach:syntaxsck}, quantifiers and forms have their standard meanings. A pattern, denoted as \texttt{pat}, is an idiom corresponding to a well-established security vulnerability in smart contracts. These patterns are identified by analyzing existing attacks
and can be easily extended.  For example, the \texttt{reentrancy} pattern on line 8 in Figure~\ref{code:votingsol} checks for the classic reentrancy vulnerability which caused the infamous DAO hack~\cite{atzei2017survey}. Even though reentrancy attacks have been largely  addressed in recent Solidity updates by limiting the gas for \texttt{send} and \texttt{transfer} functions \cite{reentrancy_solidity_soln}, the threat of other forms of reentrancy still persists as there are no gas limits for regular functions, like the \texttt{transferFrom} function in ERC777 contract on line~12-13 of Figure~\ref{code:votingsol}. Pattern \texttt{overflow} checks for integer overflow and underflow. As a word in the EVM has 256 bits, an unsigned integer faces the danger of overflow or underflow when an arithmetic operation results in a value greater than $2^{256}$ or less than $0$. 
Pattern \texttt{timestamp} check for timestamp dependencies. It is a kind of vulnerability in which the program logic depends on block timestamp, an attribute that can be manipulated by the block miner and therefore susceptible to consensus-level attacks. 
The notation \texttt{@check pat} 
checks if the given pattern \texttt{pat} is satisfied
at any point of the program.
The annotation system also supports built-in predicates for the termination status. If a function returns without errors, the \texttt{return} predicate evaluates to true and false otherwise. If a function terminates by invoking the \texttt{REVERT} opcode, the \texttt{revert} predicate evaluates to true. We describe in detail how vulnerability patterns are checked in Section~\ref{sec:generatingvc}.

When the \texttt{@coq} prefix is applied to certain annotation, \ProjName{} ports these proof obligations to Coq~\cite{barras1997coq} and launches an interactive proving session, a workflow enabled by Why3's Coq driver. 
In the example above, a property of interest would be that the mapping \texttt{stakers} always behaves as a sorted doubly linked list.
This kind of verification condition 
is hard to prove using SMT solvers and can be designated to Coq
with the \texttt{@coq} prefix. Note that assumptions, assertions, and loop invariants cannot be annotated for abstraction to Coq because they involve Hoare triples inside function bodies, which means they use intermediate variables that are mapped to compiler allocated temporaries in Yul, so for the simplicity of implementation we do not support them now.

%% file: sections/4_ir.tex
\section{Generating Annotated IR}
\label{sec:generatingir}

\begin{figure}[t]
\begin{center}
$
{\small
\begin{array}{llll}

(\textit{Lit}) & l \ \in  \ \textit{Nat} \\
(\textit{Type})  & t \ ::= \ \textit{uint256} \\
(\textit{Block}) & b \ ::= \{ \ \overline{\texttt{s}} \ \} \\

(\textit{Expr}) & e \ ::= \ \texttt{x} \ | \ \texttt{f} ( \overline{e} ) \ | \ l
\\
(\textit{Stmt}) & s \ ::= \ 
\ b \ | \ \texttt{break} \ | \ \texttt{leave} \ | \ e \ | o \\
& \ \ \ \ \ \ \ \ \ | \ \texttt{if} \ e \ b  \ | \ \texttt{let} \ \texttt{x} \ := \ e \\
& \ \ \ \ \ \ \ \ \ | \  \texttt{function} \ \texttt{f} \ \overline{\texttt{x}} \ \rightarrow \ \texttt{r} \ b \\
& \ \ \ \ \ \ \ \ \ | \  \texttt{switch} \ e \ \overline{\texttt{case} (l,  s)} \ \texttt{default} \ \overline{s} \\
& \ \ \ \ \ \ \ \ \ | \  \texttt{for} \ b \ e \ b \ b
\end{array}
}
$
\end{center}
\vspace{-5pt}
\caption{The syntax of a restricted subset of Yul}
\label{fig:mach:syntaxyul}
\end{figure}

\para{Smart contract Intermediate Representation.}
Yul~\cite{yul_docs} is an intermediate representation (IR) for Ethereum smart contracts. It is designed to be the middle layer between source languages (Solidity, Vyper, etc.) and compilation targets (EVM bytecode~\cite{wood2014ethereum}, eWASM~\cite{ewasm/design_2020}, etc.).
Yul uses the EVM instruction set, including \texttt{ADD}, \texttt{MLOAD}, and \texttt{SSTORE}, while offering native support for 
common programming constructs such as function calls, control flow statements, and switch statements. These built-in programming constructs abstract away obscure low-level instructions such as \texttt{SWAP} and \texttt{JUMP} which are difficult to reason about. As such, Yul bridges the high-level semantics written in the contract's source code and the low-level execution semantics determined by the EVM backend. Because of these advantages, we use Yul IR as the verification target for \ProjName{}. The compiling from source code to Yul is completed by 
the compiler front-end (e.g., \texttt{solc} for Solidity) 
and does not need to be trusted.
In fact, \ProjName{} can be used to detect bugs
in the compiler front-end (see Section~\ref{sec:ExperimentalEvalation}).

\para{Parsing source annotation to IR annotation} Since \ProjName{} reasons about the contract program at the Yul level,
it generates Yul-level specifications
from the source-level annotations.
Most of the annotations describing functional correctness can be straightforwardly translated (by mapping procedures and variables to their Yul-level counterparts) 
once we encoded the memory allocation mechanism of Yul. 
The challenging part is how to deal with
state variables that live in EVM storage and operate according to EVM storage rules, and data structures that live in EVM memory/storage. 

State variables in a contract are variables defined explicitly and accessible in that contract's scope. Recall from Figure~\ref{fig:mach:syntaxyul} that Yul, as an intermediate representation, does not have the notion of a state variable.
According to the Ethereum blockchain specification~\cite{wood2014ethereum}, state variables are stored in \texttt{storage}. In the backend, \texttt{storage} is manipulated by opcodes 
such as \texttt{sstore} and \texttt{sload}. For example, line 26 in Figure~\ref{code:votingsol} refers to variable \texttt{HEAD}, which, in the translated Yul file, is a pointer to a hashed location in the storage. We observe that all state variables such as \texttt{HEAD} exhibit similar behaviors at the Yul level: they are mapped to storage segments and operations on them are abstracted to the following
three functions: read ($\overline{\Psi}_r$), write ($\overline{\Psi}_w$), and metadata ($\overline{\Psi}_m$). Each operation function takes an initial identifier $id$ differentiating the location of state variables in \texttt{storage}. All three operation functions first hash the parameters and then manage EVM storage by calling opcodes \texttt{sstore} and \texttt{sload} on the hashed values. In reality, precise modeling of these hash operations and storage management is infeasible for automated SMT solvers. Therefore, we model a state variable as a tuple $\langle id, \overline{\Psi}_r, \overline{\Psi}_w, \overline{\Psi}_m \rangle$. More generically, since variable operations depend on their type (mapping, array, struct, uint256, etc.), location (storage, memory), and declaration (dynamic or static), \ProjName{} provide predefined templates of $\langle id, \overline{\Psi}_r, \overline{\Psi}_w, \overline{\Psi}_m \rangle$ tuples for each class of variables. For example, variable \texttt{past\_stakes} from line 17 in Figure~\ref{code:votingsol} is a dynamic array. Its corresponding operation tuple is shown below (here, field $id$ is \texttt{0x00} for \texttt{past\_stakes} since it is the first state variable defined in the contract): 

\begin{center}
$
{\small
\begin{array}{llll}

\textbf{id} & \texttt{0x00} \\
\overline{\Psi}_r & \texttt{fun} \ \texttt{p} \ \texttt{i} \rightarrow \texttt{read\_from\_storage\_dynamic}  \\
& \texttt{storage\_array\_index\_access\_t\_array\_storage} \ \texttt{p} \ \texttt{i}\\
 \overline{\Psi}_w & \texttt{fun} \ \texttt{p} \ \texttt{i} \ \texttt{v} \rightarrow \texttt{update\_storage\_value} \ \texttt{p} \ \texttt{i} \ \texttt{v}\\
 \overline{\Psi}_m & \texttt{fun} \ \texttt{p} \rightarrow \texttt{array\_length\_t\_array\_storage} \ \texttt{p}
 
\end{array}
}
$
\end{center}
\vspace{5pt}

When translating the annotations mentioning \texttt{past\_stakes}, \ProjName{} automatically expands abstract operators (e.g., \texttt{past\_stakes.length}) into concrete Yul function calls like $\Psi_m$ \texttt{past\_stakes}, which corresponds to \texttt{((fun p} $\rightarrow$ \texttt{array\_length\_t\_array\_storage) past\_stakes)}. Similarly, source-level specifications like \texttt{past\_stakes[0] = 0} are translated into \texttt{$\Psi_w$ $id$ 0 0}. Besides modeling storage and memory variables with abstracted functions, \ProjName{} provides additional abstraction layer refinement theorems to the proof engine to reduce the reasoning effort, while also maintaining enough detail so that layout-related specification can still be expressed and checked. For state variables in EVM storage like \texttt{past\_stakes}, we simplify operations on it by applying theorems about $\Psi_w$ and $\Psi_r$, like the following one on direct storage reduction: 
\begin{theorem}[Storage Reduce]
\[\forall \ \phi : \verb"evm_state",\ i:\verb"int",\ v:\verb"int".
\ v= \verb"pop" \ (\Psi_r (\Psi_w \ \phi \ i\ v))\]
\end{theorem}

\para{Loop invariant learning.}
\label{sec:Neural}
To verify the functional correctness of programs with loops, loop invariants must be provided. A loop invariant is a formula that holds true before and after each iteration of the loop. Decentralized gaming and finance applications, due to heavy numerical operations, often involve loops in their execution logic and require developers to generate loop invariants for verification, which is a non-trivial task. In principle, \ProjName{} can plug-in any language-independent data-driven loop invariant inference tool. Currently, \ProjName{} integrates Continuous Logic Networks (CLN)~\cite{ryan2019cln2invold,ryan2019cln2inv} into its verification workflow to automatically infer numeric loop variants based on simple user annotations. This step happens after we parsed source-level annotations and before we generate IR annotations.

Consider the \texttt{\_lotteryReward} function in Figure~\ref{code:votingsol}, which calculates the reward amount and transfers the reward to the recipient.
In order to infer the desired invariant formula, \ProjName{} requires the user to label the variable of interest using the \texttt{@learn} directive. \ProjName{} then parses the loop source code, reads the user-provided annotations, and keeps the specified variables on a list of monitored variables.

\begin{figure}[t]
\begin{center}
\includegraphics[width=\textwidth/2]{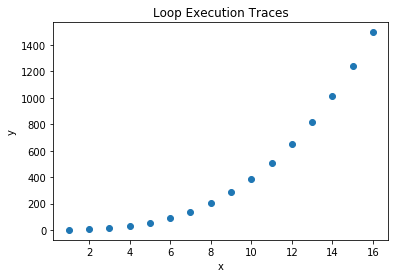}
\end{center}
\vspace{-10pt}
\caption{Traces of the while loop in the \texttt{\_lotteryReward} function}
\label{fig:traces}
\end{figure}

CLN infers invariant formulas based on loop execution traces. To obtain the traces of the \texttt{reward} function, \ProjName{} first receives a WhyML version of the smart contract IR from the translator module and injects variable monitor code which keeps track of the intermediary values of the watched variables during each iteration of the loop. \ProjName{} then executes the \texttt{reward} function based on the built-in EVM execution semantics, capturing the intermediary outputs of \texttt{x} and \texttt{y} (see Figure\ref{fig:traces}). 

The execution traces are then fed into CLN to learn the parameters and logical connectives of the formula~\cite{ryan2019cln2inv}.
For the while loop shown on line 45 in Figure~\ref{code:votingsol}, CLN learns the following invariant:
\begin{align} \label{solved_invariant}
    6x - 2y^3 - 3y^2 - y = 0.
\end{align}
The learned invariant in Eq.~\eqref{solved_invariant} is then checked by Z3 against the specifications detailed on line 42-44 in Figure~\ref{code:votingsol} and proved to be valid.

When there are sequential or nested loops in a single function, CLN learns their invariants independently by logging execution traces at each loop and training separate neural models. When CLN fails to infer the correct invariant, \ProjName{} prompts the user to manually provide a loop invariant.

%% file: sections/5_whyml.tex
\section{Translating Annotated IR into WhyML}
\label{sec:translatingirintowhyml}

\para{Intermediate verification language.} After generating the annotated
Yul IR from a smart contract source code, \ProjName{} translates the Yul IR into an intermediate verification language called WhyML (syntax in Figure~\ref{fig:mach:syntaxml})~\cite{bobot2011why3}. An intermediate verification language combines a software program with specifications and serves as a middle layer between programs and theorem provers. WhyML is a functional programming language and a part of a larger deductive verification framework Why3~\cite{filliatre2013why3}.

\begin{figure}[t]
\begin{center}
$
{\small
\begin{array}{llllllllllll}

(\textit{Lit}) &\textit{l} \ \in  \ \textit{Nat} \\
(\textit{Type})  &\textit{t} \ ::= \ int \\
(\textit{Spec}) &\textit{wsp} \ ::= \ \texttt{assume} \ \{ \ e \ \} \ | \
\texttt{assert} \ \{ \ e \ \} \\
& \ \ \ \ \ \ \ \ \ \ \ \ \ \ | \ 
\texttt{requires} \ \{ \ e \ \} \ | \
\texttt{ensures} \ \{ \ e \ \} \\
(\textit{FuncDef})  &\textit{fd} \ ::= 
\texttt{let} \ \texttt{function} \ \texttt{f} \ \textit{spec}^{*} \ = \ \textit{spec}^{*} e \\
(\textit{Expr}) &\textit{e} \ ::= \ \ 
\texttt{x} \ | \ l \ | \ e \  \oplus \ e \ | \ ( \ e \ ) \ | \ e \ ; \ e \ | \ \textit{fd} \\
& \ \ \ \ \ \ \ \ \ | \ 
\texttt{if} \ e \ \texttt{then} \ e \\
& \ \ \ \ \ \ \ \ \ | \ 
\texttt{while} \ e  \ \texttt{do} \ \textit{wsp}^{*} \ e \ \texttt{done} \\
& \ \ \ \ \ \ \ \ \ | \ 
\texttt{let} \ \texttt{x} \ = \ e \ \texttt{in} \ e \\
& \ \ \ \ \ \ \ \ \ | \ 
\texttt{x} \ := \ e \\
& \ \ \ \ \ \ \ \ \ | \ 
\texttt{match} \ e \ \texttt{with} \ \overline{( | \ \textit{pattern} \ \rightarrow \ e)}^{+} \ \texttt{end} \\
& \ \ \ \ \ \ \ \ \ | \ 
e \ e^{+} \ | \ \{ \ \overline{\texttt{x} = \texttt{e}} \ \} \ | \ e.\texttt{x} \ | \ \texttt{raise} \ \texttt{x} \\
& \ \ \ \ \ \ \ \ \ | \ 
e \ : \ t \ | \ e \ \textit{wsp}^{+} \\
& \ \ \ \ \ \ \ \ \ | \ 
\texttt{try} \ e \ \texttt{with} \ \texttt{x} \rightarrow e \ \texttt{end} \\
& \ \ \ \ \ \ \ \ \ | \ 
\texttt{begin} \ e \ \texttt{end}\\
& \ \ \ \ \ \ \ \ \ | \ 
\texttt{function} \ \overline{\texttt{x}} \ \rightarrow \ e \\

(\textit{Exception}) &\textit{exp} \ ::= \texttt{exception} \ \texttt{x} \\
(\textit{Decl}) &\textit{wsp} \ ::= \ \textit{exp} \ | \ \textit{fd} \\
(\textit{Module})  &\textit{module} \ ::= \
\texttt{module} \ \texttt{m} \ \textit{decl}^* \ \texttt{end}\\
(\textit{Prog})  &\textit{p} \ ::= \ \overline{(\textit{theory} \ | \ \textit{module})}^{*}\\

\end{array}
}
$
\end{center}
\caption{The syntax of a restricted subset of WhyML.}
\label{fig:mach:syntaxml}
\end{figure}

\para{EVM execution semantics modeling.}
Since Yul embeds EVM opcodes, we provide an EVM formalization as a WhyML library. Our library formalizes opcode semantics as a state machine using function primitives. Specifically, EVM operations in Yul can either be pure functions (i.e., functions that do not generate side-effects, such as \texttt{ADD}, which simply returns the sum of the two input values) or impure (such as \texttt{SSTORE}, which modifies the underlying storage of the EVM). The presence of both pure and impure functions in Yul not only makes it difficult to translate Yul to WhyML, but also poses a challenge in analyzing EVM state during the execution of a function. To address these two issues, we make our EVM model functional, which means that every EVM instruction takes an input state and returns an output state. Impure functions can directly return a state with updated attributes (e.g., \texttt{SSTORE} returns a new EVM state with updated storage) and pure functions simply return a state with an updated \texttt{stack}, which holds the results of these computations.  We extend prior EVM formalizations, including a Lem formalization~\cite{hirai_pirapira/eth-isabelle_2020} and Imandra EVM \cite{imandra_aestheticintegration/contracts}, and present the first implementation in WhyML. In our execution semantics model, we specify the semantics of the EVM state, message, memory, storage, and instructions, according to the Ethereum Yellow Paper~\cite{wood2014ethereum}, which outlines standard EVM behaviors.

We now walk through two components of \ProjName{}'s EVM execution semantics: the EVM state and the memory model. Since EVM is a stack-based machine with a word size of 256-bit, we introduce the \texttt{word} type using WhyML's built-in integer theory and place a bound $0 \leqslant i < 2^{256}$ to model the valid range of the bit array. 

\begin{figure}[t]
\begin{center}
\begin{lstlisting}[language = Why3]
type array_t   = { data       : list word;
                   size       : word       }
                    
type message_t = { recipient  : word;
                   sender     : word;
                   value      : word;
                   gas        : word       }
  
type state_t   = { stack      : list word; (* stores return values *)
                   calldata   : list word; (* stores input params *)
                   memory     : array_t;
                   storage    : array_t;
                   message    : message_t;
                   pc         : word;      }
\end{lstlisting}
\end{center}
\vspace{-10pt}
\caption{EVM state definition}
\label{fig:evm_state}
\end{figure}

We define the EVM state in Figure~\ref{fig:evm_state}. EVM's memory is a storage of temporary data that persists during smart contract execution and is deleted at the end of the contract call. Since EVM memory is a word-addressed byte array, we implement the \texttt{data} field of the memory with a list of words using the built-in list theory in WhyML. The \texttt{size} field stores the current number of bytes in the memory. 

In Yul IR, EVM operations can either be pure functions (i.e., functions that do not generate side-effects, such as \texttt{ADD}, which simply returns the sum of the two input values) or impure (such as \texttt{SSTORE}, which modifies the underlying storage of the EVM). The presence of both pure and impure operations in Yul not only makes it difficult to translate Yul to WhyML, but also poses a challenge in analyzing EVM state during the execution of a function. To address this issue, we make our EVM model functional, which means that every EVM instruction takes an input state and returns an output state. Impure functions can directly return a state with updated attributes (for example, \texttt{SSTORE} returns a new EVM state with an updated storage), and pure functions simply return a state with an updated \texttt{stack}, which holds the results of these computations. 

We present an example of \ProjName{}'s EVM instruction semantics modeling in Figure~\ref{fig:mload}. The \texttt{MLOAD} instruction takes an EVM state as input and outputs a new EVM state. 
We implement functions such as \texttt{get\_mem} and \texttt{pad\_mem} which simulate the internal behaviors of the memory. 
According to the Ethereum Yellow Paper, EVM memory automatically adjusts its size when the memory location being accessed is greater than the current memory size, under which circumstance it resizes itself to the queried location and dynamically increases the gas fee.

\begin{figure}[t]
\begin{center}
\begin{lstlisting}[language = Why3]
(* Extract param from state *)
let ghost function param (s : State.t) (i : int) : word
= let params = s.cd in nth i params

(* EVM internal operations *)
let function ceiling (x : word) : word
= (x + byte_size) / byte_size

let function round_up (max_index i : word) : word
= let end_idx = i + byte_size in max max_index (ceiling end_idx)
    
let rec function k_zeroes (k : word) : list word
variant { k }
= if k < 0 then (Nil : list word) 
  else (Cons 0 (Nil : list word)) ++ (k_zeroes (k - 1))

let function pad_mem (data : list word) (cur_size new_size : word) : list word
= let d = data in let zeros = k_zeroes (new_size - cur_size) in d ++ zeros

let ghost function get_mem (byte_idx : word) (mem : t) : t
= let size_w = mem.cur_size in
  if byte_idx < (size_w * 32) then
    let d = nth (byte_idx / 32) mem.data in { mem with peek = d }
  else
    let new_size = round_up mem.cur_size byte_idx in
    let new_data = pad_mem mem.data mem.cur_size new_size in
    { data = new_data; cur_size = new_size; peek = 0 }

(* MLOAD instruction *)
let ghost function mload (s : state_t) : state_t
= let mem' = (get_mem (param s 0) s.m) in 
  { s with m = mem'; sk = Cons mem'.peek Nil }
\end{lstlisting}
\end{center}
\vspace{-10pt}
\caption{Execution semantics of the \texttt{MLOAD} instruction in WhyML}
\label{fig:mload}
\end{figure}

\begin{figure}[t]
\begin{center}
\begin{lstlisting}[language = Why3]
(* s0 is the input state. s1 is the output state. *)
(* Gas cost of MLOAD is proportional to memory resource consumption. *)
assert { let cost = (mem_cost s1.m - mem_cost (s0.m) + gas_verylow) in
         s1.gas = (s0.gas) - cost }

(* If queried memory address exceeds memory bound, extend the address size such that the address is included. *)
assert { (byte_idx < (s0.m.size) /\ s1.m.size = (s0.m.size)) 
         \/ (byte_idx >= (s0.m.size) 
             /\ s1.m.size = (word_ceiling (s0.m.size) byte_idx)) }

(* Content of memory remains the same after MLOAD. *)
assert { forall i. 0 <= i <= s1.m.size -> s1.m.data[i] = old s1.m.data[i] }
\end{lstlisting}
\end{center}
\vspace{-10pt}
\caption{Specifications of \texttt{MLOAD} according to Ethereum Yellow Paper}
\label{fig:mem_spec}
\end{figure}
To ensure our formalized EVM model is correct, we write specifications for the opcode instructions and then verify their correctness against our implementation in WhyML. Take the \texttt{MLOAD} function for example, the Yellow Paper describes the following behaviors of \texttt{MLOAD}:
\begin{itemize}
    \item The gas fee associated with \texttt{MLOAD} is proportional to the smallest multiple of 32 bytes that can include all memory indices in its range;
    \item If the queried memory address exceeds the current bound of active memory, memory is extended to include the queried memory address;
    \item The content of the memory, excluding extended sections, does not change after each \texttt{MLOAD}.
\end{itemize}
The specifications of \texttt{MLOAD} are discharged by the translator as \texttt{assert} statements in the translated WhyML contract, as shown in Figure~\ref{fig:mem_spec}. With the opcode instructions checked against the expected behaviors outlined in the Yellow Paper, we ensure that the EVM model conforms to the formal semantics and provides the capability to reason about the correctness of smart contracts from a low-level view.

\para{Translating Yul IR to WhyML.} With the EVM model implemented in WhyML, we can now translate the Yul program into WhyML. The translation rules are defined in Figure~\ref{fig:translation}. We use $\overline{e}$ to denote a sequence of expressions, with the expanded form of $e_1, e_2, ... , e_i$. We write $e \triangleright \kappa$ to denote that expression $e$ can be translated into $\kappa$ in \ProjName{}. $\Sigma$, the exception identifier, and its return value symbol, $\sigma$ of type $unit$, are used to mimic imperative control flow. $\pi$ denotes the variable to be returned (corresponding to \texttt{r} in the Yul syntax in Figure\ref{fig:mach:syntaxyul}), and $\phi$ denotes the EVM state variable that we explicitly pass throughout instructions.

\begin{figure*}[t]
    \begin{mathpar}
\inferrule*[Right=(Var)]{  }{ \vdash \texttt{x} \triangleright \texttt{x} }

\and
\inferrule*[Right=(VarRet)]{{ \vdash IsReturnVar(\texttt{x})}
}{ \vdash \texttt{x} \triangleright \pi }

\and

\inferrule*[Right=(Leave)]{ \texttt{leave} }{ \texttt{raise} \ \Sigma_l }

\\

\inferrule*[Right=(Call)]{{ \vdash \overline{e} \triangleright \overline{\kappa}}\\
{ \vdash \texttt{f} \triangleright \Psi}}{ \vdash \texttt{f} \ ( \ \overline{e} \ ) \ \triangleright \ ( \ \phi \ := \ \texttt{push} \ \overline{\kappa} \ ; \phi \ := \ \Psi \ \phi \ ; \ \texttt{pop} \ \phi \ ) }

\and 
\inferrule*[Right=(If)]{
{\vdash e \triangleright \kappa}\\
{\vdash b \triangleright \overline{\tau} }}
{ \vdash \texttt{if} \ e \ b \ \triangleright \ ( \ \texttt{if} \ \kappa \ \texttt{then} \  \overline{\tau}\ ) \ ; }

\\

\inferrule*[Right=(Asg)]{
{\vdash \texttt{x} \triangleright \texttt{x}}\\
{\vdash e \triangleright \kappa } }{ \vdash \texttt{let} \  \texttt{x} := e \ \triangleright \ \texttt{let} \ \texttt{x} \ = \kappa \ ;}
\and

\inferrule*[Right=(Seq)]{
{\vdash \texttt{x} \triangleright \texttt{x}}\\
{\vdash e_1 \triangleright \kappa_1}\\
{\vdash e_2 \triangleright \kappa_2}
 }{ \vdash \texttt{let} \ \texttt{x} := e_1 \ , \ e_2 \ \triangleright \ \texttt{let} \ \texttt{x} \ = \kappa_1 \ \texttt{in} \ \kappa_2}
 
\\

\inferrule*[Right=(Switch)]{
{\vdash e \triangleright \kappa}\\
{\vdash \overline{(l , s)} \triangleright \overline{(\ell, \tau)}}\\
{\vdash \overline{s}_i \triangleright \overline{\tau}}_i\\
}{ \vdash \texttt{switch} \ e \ \overline{\texttt{case} \ (l, s)} \ \texttt{default} \ \overline{s}_i \ \triangleright \ \texttt{match} \ \kappa \ \texttt{with} \ \overline{( \ | \ \ell \ \rightarrow \ \tau \ )} \ \overline{\tau}_i \ \texttt{end}}

\\

\inferrule*[Right=(Loop)]{
{\vdash e \triangleright \kappa }\\
{\vdash b_1 \triangleright \overline{\tau}_1}\\
{\vdash b_2 \triangleright \overline{\tau}_2}\\
{\vdash b_3 \triangleright \overline{\tau}_3}\\
}{ \vdash \texttt{for} \ b_1 \ e \ b_2 \ b_3 \ \triangleright \ \overline{\tau}_1 \ ; \ \texttt{while} \ \kappa \ \texttt{do} \ ( \ \texttt{try} \ \overline{\tau}_2 \ ; \ \overline{\tau}_3 \ \texttt{with} \ \Sigma \rightarrow \sigma \ \texttt{end} \ ) \ \texttt{done} }

\\

\inferrule*[Right=(Def)]{
{\vdash \texttt{f} \triangleright \Psi}\\
{\vdash \texttt{r} \triangleright \pi}\\
{\vdash \overline{\texttt{x}} \triangleright \overline{\kappa}}\\
{\vdash b \triangleright \overline{\tau}}}
{ \vdash \texttt{function} \ \texttt{f} \ \overline{\texttt{x}} \ \rightarrow \texttt{r} \ b \ \triangleright \\ \ \texttt{let} \ \texttt{function} \ \Psi \ \overline{\kappa}
\ = 
\ ( \ 
\texttt{try} \ 
\overline{\tau}
\ \texttt{with} \ \Sigma_l \rightarrow \texttt{push} \ \pi \ \texttt{end}
\ ) \ )
}
\end{mathpar}
    \vspace{-15pt}
    \caption{The translation inference rules of \ProjName{}.}
    \label{fig:translation}
\end{figure*}

We present the inference rule for translating Yul IR to WhyML in Figure~\ref{fig:translation}. We now expand on how \ProjName{} treats certain Yul language features.
Ethereum smart contracts provide globally available variables about the current transaction and block, accessible by direct reference at the source level (front-end languages). For example, \texttt{msg.sender} returns the address of the function caller, \texttt{msg.value} returns the amount of wei (the smallest denomination of ether) wired to current contract.

When translating the contract program, \ProjName{} treats these opcodes just like function calls. This is enabled by the aforementioned EVM model, which defines concrete behaviors for the opcodes. As shown in Figure~\ref{code:votingsol}, the \texttt{msg.sender} on line 10 is translated into \texttt{caller()}. On the other hand, the translation of certain opcode calls written in user-provided annotations
is treated differently. For example, the \texttt{msg.sender} on line 5, which is being used in the annotation rather than in the program, is translated into \texttt{$\phi$.message.sender}.

Dealing with the complex semantics of inter-contract calls is also challenging.
In particular, we specify functions being Effectively Callback Free (ECF)~\cite{grossman2017online}. 
ECF is a property that avoids most blockchain-level bugs by restricting R/W on transaction states (e.g., globally available variables and storage) after calling external contract functions. Existing algorithms for detecting ECF either are online or inspect deployed bytecode and transaction histories. Since it is impossible to know what other contracts do before deployment, \ProjName{} requires input from the developer to secure the ECF property. 
It asks if a function $f$ changes the EVM state.
If it is, \ProjName{} proceeds by adding an $\texttt{axiom}$ specifying that the return state of $f$ satisfies the same predicate as if there were no callback altering the original $\phi$. 
For example, if we have an external contract call like \texttt{transferFrom} on line~12-13 of Figure~\ref{code:votingsol}, \ProjName{} inserts
an \texttt{axiom} shown in Figure~\ref{fig:ecf},
where \texttt{push} is used to load the two arguments of \texttt{sendEtherTo} into state \texttt{st}. 

\begin{figure}[t]
\begin{center}
\begin{lstlisting}[language = Why3]
axiom transferFromECF : forall st: evmState, st': evmState, sender: address, receiver: address, amount: uint256.
    st = push sender receiver amount
    -> st' = transferFrom st
    -> st' = st
\end{lstlisting}
\end{center}
\vspace{-10pt}
\caption{WhyML axiom of \texttt{transferFrom}'s ECF condition}
\label{fig:ecf}
\end{figure}


For state variables and data structures, \ProjName{} maps them to abstract data types in $\phi$. For example, operations on the \texttt{userVoted} state variable on line 17 in Figure~\ref{code:votingsol}, which was translated to a set of $\Psi_w$, $\Psi_r$, and $\Psi_m$ on function ids in the last phase, are further translated into abstract map operations, namely, uninterpreted functions. As shown by step 2 and step 3 in Section~\ref{sec:Overview}, \texttt{$\Psi_r$ 0x05 i}, reading element at \texttt{i} of mapping \texttt{userVoted}, will be translated into \texttt{Map.get $\phi$.map\_0x05 i}.

Yul has the traditional C-style implicit type conversions between values of type \texttt{uint256} and \texttt{bool}. Since WhyML does not support such features, \ProjName{} simulates this implicit conversion by restricting the type of Yul to only \texttt{int} and treats all operations on \texttt{bool} with \texttt{iszero}, with explicit conversion functions. The reason we do not model our types using 256-bit bitvectors is that its theory is not scalable with certain SMT solvers. With integers, reasoning can be much simplified. 

There exists a gap (scoping, continuation, and return type) between Yul, an imperative IR, and WhyML, a functional language. \ProjName{} fills this gap by mimicking imperative control flow constructs (e.g., \texttt{while}, \texttt{for}, \texttt{return}, and \texttt{break}) by exceptions. Since Yul only has well-structured control flow operators (i.e., no \texttt{goto}), this modeling is sound in the sense that it creates the exact same CFG structure. 

We also experimented with using a fully monadic translation style, where we model all the imperative continuations and mutations through monad transformers, but our experiments showed that this representation actually made the SMT clauses much more complex (through the higher-order reasoning of \texttt{unit} and \texttt{bind}) and therefore cannot be efficiently solved. More importantly, a monadic representation of the program breaks the specification written at the front-end level. For example, loop invariants cannot be easily translated to their counterparts, and often we have to manually rewrite the specifications at the IR (Yul) level.

%% file: sections/6_vc.tex
\section{Generating Verification Conditions}
\label{sec:generatingvc}

\ProjName{} generates a variety of verification conditions at different abstraction levels from WhyML. With various drivers provided by Why3, \ProjName{} can leverage SMT solvers as well as the Coq proof assistant to verify the correctness specifications.

\para{Checking vulnerability patterns.} 
The Ethereum community has long observed security attack patterns and set up community guidelines to avoid pitfalls during contract development. Recall that \ProjName{} allows the user to specify a \texttt{@check} directive at the source level to detect certain pre-defined vulnerability patterns (see Figure~\ref{fig:mach:syntaxsck}). \ProjName{} implements the verification 
for three vulnerability patterns: integer overflow, reentrancy, and timestamp dependence.
\ProjName{} uses WhyML and static analysis to perform pattern checking. For the \texttt{overflow} pattern, \ProjName{} generates specifications in WhyML to be verified. For the \texttt{reentrancy} and \texttt{timestamp} patterns, \ProjName{} performs static analysis to detect any potential vulnerabilities.

If the \texttt{@check overflow} annotation is specified, \ProjName{} automatically adds assertions to all the integer variables that check whether the program handles integer overflow and underflow. For an unsigned integer variable, $v$, that has $b$ bits in size, WhyML inserts the following assertion at the end of the function:
\[
    \texttt{assert} \, \{ \, \neg (0 \leqslant v < 2^{b}) \Rightarrow \texttt{revert} \, \} \quad \text{(unsigned integer)}
\]
Similarly, for a signed integer variable, WhyML inserts the following assertion at the end of the function:
\[
    \texttt{assert} \, \{ \, \neg (-2^{b/2} \leqslant v < 2^{b/2}) \Rightarrow \texttt{revert} \, \} \quad \text{(signed integer)}
\]
where \texttt{revert} is the termination status in the annotation system.
We choose to place the assertions at the end of a function, instead of placing right after arithmetic operations. This allows for correct checking of the following Ethereum smart contract development common practice: developers first intentionally perform an unsound arithmetic operation and then check if the result is larger (in case of underflow) or smaller (in case of overflow) than the two operands. If it is, revert the EVM state (i.e., exit the program and rollback to previous memory and storage state), otherwise continue program execution. In this case, using the naive overflow check (placing assertion immediately after variable definition) will give a false warning on the intentionally overflowed (underflowed) variable. However, with our approach, intentionally overflowed variables will have enough time to propagate their results to where smart contract developers manually check overflow. We notice that \texttt{solc-verify}~\cite{hajdu2019solc} also adopt this approach. However, since their annotation system and framework is at the source (Solidity) level, they introduce extra false negatives by delaying the assertion. Consider a overflowed variable \texttt{x}, the delayed assertion at the source code-level permits an edge case where \texttt{x} is modified by operations before the delayed assertions such that it still passes certain overflow checks. Now the approach using delayed assertions at the source-code level does not detect the overflow incident. In our approach, since \ProjName{} discharges VCs at the IR level (where every variable is in SSA form) and Yul creates temporary variables for assignments, the reassignment of \texttt{x} at the end of the function can detect the overflow incident happened to a snapshot of the \texttt{x} variable.

This approach also has its limitation when certain SMT solvers cannot handle 256-bit integers; however, smart contract development has focused increasingly on gas efficiency, which promotes the use of cheaper data types~\cite{gashigh} such as \texttt{uint64}, which can be handled by most SMT solvers.

To detect time dependency vulnerabilities, \ProjName{} adds specifications requiring that variables within the contract do not depend on the current timestamp, which can be manipulated by the miner~\cite{known_attacks}.
To detect the \texttt{reentrancy} pattern, \ProjName{} performs a static analysis on the Yul IR level to ensure that all storage related-operations are placed before external contract calls. If this property is violated, \ProjName{} produces a warning signaling a potential reentrancy error. 

\para{Discharging SMT solvers.}
For each of the verification condition pair (pre-condition and post-condition), \ProjName{} discharges VCs for a SMT solver chosen by the user, a process enabled by Why3's drivers. Currently, the supported solvers are Z3~\cite{TACAS:deMB08}, Alt-Ergo~\cite{bobot2013alt}, and CVC4~\cite{barrett2011cvc4}.

As mentioned in the prior section, some SMT solvers are not capable of handling 256-bit unsigned integers. Without loss of generality, we reduce the word size to 64 bits in Figure~\ref{fig:arr_overflow_ir} and Figure~\ref{fig:arr_overflow_whyml} when demonstrating the vulnerability.

\para{Discharging Coq proof assistant.}
For annotations written with \texttt{@coq}, \ProjName{} discharges the Coq proof assistant by directly calling WhyML's porting mechanism, which shallowly embeds annotated WhyML program as Gallina, Coq's specification language. For example, the ported Coq proof obligation for annotation on line 21 and 22 in Figure~\ref{code:votingsol} is as follows:
\begin{lstlisting}[language = Coq,basicstyle=\small]
Theorem rebalanceStakers'vc :
  sorted_doubly_linked_list (map_0x03 st_c) 
  -> let st_c2 := rebalanceStakers st_c in
     sorted_doubly_linked_list (map_0x03 st_c2).
\end{lstlisting}
In the above theorem, \texttt{st\_c} is the EVM state when user is calling the function \texttt{rebalanceStakers}. It has a Coq \texttt{Record} type with a field \texttt{map\_0x03} defined as an uninterpreted function from addresses to addresses. \texttt{map\_0x03} is an abstraction of the source-level variable \texttt{stakers}, which stores the addresses of users who participated in the staked voting. The \texttt{st\_c} variable is generated as a \textit{Parameter} (bound to its type in the global environment) in the Coq file. The \texttt{sorted\_doubly\_linked\_list} is a user defined property which asserts that the mapping forms a sorted doubly linked list. \texttt{rebalanceStakers} is also defined as an uninterpreted function with its behavior bounded by an axiom in the environment called \texttt{rebalanceStakers'def} which embeds the function body in Gallina.

%% file: sections/7_case_study.tex
\section{Evaluation}
\label{sec:ExperimentalEvalation}

\para{Environment.} All the experiments are conducted on a Ubuntu 18.04 system with Intel Core i7-6500U @ 4x 3.1GHz and 8GB RAM. The specific versions of tools used are Solidity 0.8.1, Why3 1.3.1, Alt-ergo 2.3.2, Z3 4.8.6, and CVC4 version 1.7. We set the timeout of SMT solvers in \ProjName{} to 5 seconds.

\para{Benchmarks.} 
We compiled a set of smart contracts, a part from prior work on VeriSol~\cite{wang2019formal}, that reflects common uses of Solidity, and a real-world Decentralized Finance (DeFi)~\cite{chen2020blockchain} protocol contract that mimics Uniswap~\cite{uniswap_whitepaper}. We show the details of those benchmarks in Table~\ref{tab:libraryspec}.  The ``Scenario'' column briefly explains each contract's purpose. For example, the contract \texttt{BazaarItemListing} allows users to list items publicly on the blockchain and then trade in a decentralized way; therefore its scenario is \textit{Market}. The \textit{Governance} scenario means that the contract is used for managing interactions between multiple on-chain parties. Other scenarios like \textit{Game} and \textit{Supply Chain} means the contract encodes application-specific business logic.
The ``LOC'' column is defined as the total lines of Solidity program, and the length of the generated Yul program is shown in the ``LOC (Yul)'' column.

\begin{table}[t]
   \caption{\ProjName{} benchmark statistics.}
    \centering
    {\begin{tabular}{lrrr}
        \toprule
        Contract & Scenario & LOC & LOC (Yul) \\
        \midrule
        AssetTransfer~\cite{wang2019formal} & Banking & 218 & 1,789 \\
        BasicProvenance~\cite{wang2019formal} & Governance & 53 & 686 \\
        BazaarItemListing~\cite{wang2019formal} & Market & 130 & 3,653 \\
        DefectiveComponentCounter~\cite{wang2019formal} & Utility & 44 & 1,026 \\
        DigitalLocker~\cite{wang2019formal} & Utility & 152 & 1,636 \\
        FrequentFlyerRewardsCalculator~\cite{wang2019formal} & Utility & 60 & 1,072 \\
        HelloBlockchain~\cite{wang2019formal} & Education & 45 & 1,013 \\
        PingPongGame~\cite{wang2019formal} & Game & 95 & 2,964 \\
        RefrigeratedTransportation~\cite{wang2019formal} & Supply Chain & 145 & 1,865 \\
        RefrigeratedTransportationWithTime~\cite{wang2019formal} & Supply Chain & 110 & 1,912 \\
        RoomThermostat~\cite{wang2019formal} & Utility & 50 & 755 \\
        SimpleMarketplace~\cite{wang2019formal} & Exchange & 74 & 1,055 \\
        UniswapStyleMarketMaker~\cite{uniswap_whitepaper} & Exchange & 135 & 2,197 \\
        \bottomrule
    \end{tabular}}
    \label{tab:libraryspec}
\end{table}

\para{Properties of interest.} 
To understand the common
types of desired security properties 
in real-world smart contracts,
we conduct a thorough analysis of our smart contract benchmark and 
the smart contracts of
167 real-world projects audited by CertiK
from the year of 2018 to 2020~\cite{ctk2020},
including 
19 DeFi protocols whose TVL (total value locked) are
larger than 20M USD,
and categorize six types of desired security properties for smart contracts.
\vspace{10px}
\begin{enumerate}
    \item[T1:] \label{prop:1} \emph{User-defined functional properties} 
    represent whether smart contract implementations satisfy the developer's intent.
    \item[T2:] \label{prop:2} \emph{Access control requirements} define proper authorization of users to execute certain functions in smart contracts.
    \item[T3:] \label{prop:3} \emph{Contract-level invariants} define properties of the global state of smart contracts that hold  at all times. 
    \item[T4:] \label{prop:4} \emph{Virtual-machine-level properties} specify the low-level behaviors of state machine to ensure the execution environment functions as expected.
    \item[T5:] \label{prop:5} \emph{Security-pattern-based properties} require that the smart contract does not contain insecure patterns at the IR level.
    \item[T6:] \label{prop:6}
    \emph{Financial model properties} define the soundness of economic models in decentralized financial systems.
\end{enumerate}

\begin{table*}
  \caption{Experimental results of \ProjName{}}
  \centering
  \begin{threeparttable}
  \footnotesize
  \begin{tabular}{lrrrrrrrrr}
    Contract & T1 & T2 & T3 & T4 & T5 & T6 & Total & Solved & Time(s) \\
        \midrule
        AssetTransfer~\cite{wang2019formal} & 10 & 19 & 0 & 2 & 3 & 0 & 34 & 32 & 8.86 \\
        BasicProvenance~\cite{wang2019formal} & 2 & 2 & 1 & 0 & 0 & 0 & 5 & 5 & 4.84 \\
        BazaarItemListing~\cite{wang2019formal} & 5 & 3 & 0 & 1 & 3 & 0 & 12 & 12 & 8.95 \\
        DefectiveComponentCounter~\cite{wang2019formal} & 2 & 1 & 0 & 2 & 2 & 0 & 7 & 7 & 4.09 \\
        DigitalLocker~\cite{wang2019formal} & 10 & 10 & 1 & 7 & 0 & 0 & 28 & 27 & 13.57 \\
        FrequentFlyerRewardsCalculator~\cite{wang2019formal} & 2 & 1 & 0 & 2 & 2 & 0 & 7 & 5 & 3.69\tnote{*} \\
        HelloBlockchain~\cite{wang2019formal} & 2 & 1 & 1 & 0 & 0 & 0 & 4 & 4 & 2.25 \\
        PingPongGame~\cite{wang2019formal} & 5 & 0 & 0 & 1 & 3 & 0 & 9 & 9 & 5.26 \\
        RefrigeratedTransportation~\cite{wang2019formal} & 3 & 10 & 0 & 0 & 3 & 0 & 16 & 16 & 6.29 \\
        RefrigeratedTransportationWithTime~\cite{wang2019formal} & 3 & 3 & 0 & 0 & 6 & 0 & 12 & 12 & 7.70 \\
        RoomThermostat~\cite{wang2019formal} & 3 & 3 & 1 & 0 & 1 & 0 & 8 & 8 & 2.95 \\
        SimpleMarketplace~\cite{wang2019formal} & 3 & 6 & 1 & 1 & 1 & 0 & 12 & 12 & 5.89 \\
        UniswapStyleMarketMaker~\cite{uniswap_whitepaper} & 0 & 0 & 0 & 1 & 0 & 3 & 4 & 4 & 1.25\tnote{*} \\
        \bottomrule
        Total & 50 & 59 & 5 & 17 & 24 & 3 & 158 & 153 & 75.59 \\
        Average time for each property (s) & 0.75 & 0.18 & 1.54 & 1.11 & 0.05 & 0.13 & - & - & - \\
        \bottomrule
  \end{tabular}
  \begin{tablenotes}
  \item[*] The total verification time shown here is only for properties that can be automatically solved by Z3. The verification of both contracts also include a manual proof of around 200 lines of Coq, whose time is not included in the table.
  \end{tablenotes}
  \end{threeparttable}
  \label{tab:results}
\end{table*}

\para{Results.} \ProjName{} successfully verifies a total of 153/158 security properties listed in Table~\ref{tab:results}. 
The security properties are categorized into the six types above. In the table, T1 to T6 correspond to the six types of security properties. 
The ``Time(s)'' column shows the total time for \ProjName{} to prove all six types of security properties. 
Out of the 153 verified properties, 136 are verified at the WhyML level via SMT solvers, two are verified by porting definitions to Coq, and 15 are verified using static analysis. 

Now we investigate the reasons why certain checks have failed. 
Verification conditions in \ProjName{} fail when SMT solvers either timeout or return a counterexample.

In our experiment, out of the five properties that \ProjName{} failed to verify, all were due to timeout. 
After inspecting the generated VCs, we found that there are two reasons for solver timeout. First, some of the translated contract representations in WhyML have too many, often redundant, verification conditions.
In these cases, we manually compress the WhyML program by removing temporary variables to reduce redundancy. Second, certain VCs are inherently complex due to the logic of the contract as well as user annotations. We find that adding intermediary assertions and using Why3's goal-splitting transformations can help the SMT solvers terminate. 
\ProjName{} successfully verifies all five of the manually retrofitted version of the WhyML contracts. Given that Yul has a stable formal semantics on which the Solidity compiler is implemented, we plan to automate
such a retrofitting process while respecting the official Yul semantics in future work.

\begin{figure}[t]
\begin{center}
\begin{lstlisting}[style=solidity]
/* @meta _token0.balanceOf(address(this)) = reserve0
 * @meta _token1.balanceOf(address(this)) = reserve1 */
contract AMM {
    ...
    uint reserve0;
    uint reserve1;

    /* @pre (token0.allowance(fromA, address(this)) == 1);
     * @coq @post (reserve0 * reserve1 > old reserve0 * old reserve1)
     * @post (token0.balanceOf(fromA) < old token0.balanceOf(fromA))
     * @post (token1.balanceOf(fromA) > old token1.balanceOf(fromA)) */
    function trade (address fromA, uint amount0) returns (uint) {
        require (amount0 > 0);
        require (fromA != address(this));
        require (reserve0 > 0);
        require (reserve1 > 0);
        bool success = _token0.transferFrom (fromA, address(this), amount0);
        assert (success);
        uint swapped = swap(fromA);
        return swapped;
    }

    function swap0to1 (address toA) returns (uint) {
        uint balance0 = _token0.balanceOf(address(this));
        uint amount0In = balance0 - reserve0;
        assert (toA != token0 && toA != token1);
        assert (amount0In > 0);
        assert (reserve0 > 0 && reserve1 > 0);
        uint amountInWithFee = amount0In * 997;
        uint numerator = amountInWithFee * reserve1;
        uint denominator = reserve0 * 1000 + amountInWithFee;
        uint result = numerator / denominator;
        bool success = _token1.transfer(toA, result);
        assert (success);
        reserve0 = _token0.balanceOf(address(this));
        reserve1 = _token1.balanceOf(address(this));
        return result;
    }
    ...
}
\end{lstlisting}
\end{center}
\vspace{-15pt}
\caption{Simplified automated market maker contract.}
\label{code:ammsol}
\end{figure}

\vspace{10px}
\para{Case study on T6 properties.} To demonstrate
that  \ProjName{} can verify
the correctness of decentralized financial models
for smart contracts, 
we study a Uniswap-style~\cite{angeris2019analysis} constant product market maker contract (135 LOC in Solidity). It represents the standard practice adopted by the constant product market makers in the current DeFi ecosystem. Uniswap and its variants are the biggest liquidity source for decentralized cryptocurrency trading (at the time of writing, they collectively carry 6B USD token market capital and a 5B USD total value locked). These contracts have various desirable properties, e.g., resistance to price manipulation. However, these properties were only proven on paper, and in fact some of the underlying assumptions could be easily violated in practice, leading to high-profile price manipulation attacks that stole millions worth of cryptocurrencies from the contract users (e.g., the bZx protocol hack on February 18, 2020~\cite{bzx2020}). \ProjName{} is able to prove the financial security properties against the contract implementation and avoid such attacks. We show an example contract in Figure~\ref{code:ammsol} where \ProjName{} proves its soundness by verifying three lemmas:

\begin{itemize}
    \item $\mathcal{L}_1$ (\textit{Swap correctness}): if one calls the \texttt{swap0to1} function, he will lose some \texttt{token0} in return for \texttt{token1}, and vice-versa.
    \item $\mathcal{L}_2$ (\textit{Reserve synchronization}): the \texttt{reserve} variable in the market maker contract always equals the actual liquidity it provides, i.e., its token balances.
    \item $\mathcal{L}_3$ (\textit{Increasing product}): constant product market makers are named because the product of the contract's reserves is always the same before and after each trade. But in our example, due to the 0.3\% transaction fee, the product no longer stays constant but increases slightly after each trade. 
\end{itemize}

The WhyML program generated from the Yul IR consists of 2,197 LOC. We first tried to prove the relatively simple property $\mathcal{L}_1$, but, due to the length of the program, the solvers timed out. After a manual inspection into the code, we discovered that it has too many low-level typecasts and temporary variables that overburdened the SMT solvers. Then we tried to manually prove these theorems, but found that the generated raw Coq file was too large (345K LOC) to reason about. In short, the complexity of DeFi has rendered reasoning about low-level programs difficult.

\begin{figure}[t]
\begin{center}
\begin{lstlisting}[language = Why3]
  let ghost function fun_swap0to1_552 (st_c: evmState)  (vloc_toA_444: int) 
      : evmState
      requires { st_c.var_0x07 
                 = pop (fun_balanceOf_33 (push st_c st_c.this.address)) }
      ...
  = let _r: ref int = ref 0 in
    let ghost st_g: ref evmState = ref st_c in
    let bvconstzero = 0 in
    try
      begin
        ... the trading procedure ...
        st_g := { st_g with var_0x07 = pop (!st_g) };
        let _51 = vloc_result_512 in
        let expr_549 = _51 in
        _r := expr_549;
        leave
      end;
      st_g := (setRet !st_g !_r);
      raise Ret
    with Ret -> (!st_g) 
    end
\end{lstlisting}
\end{center}
\vspace{-10pt}
\caption{Snippet of the retrofitted \texttt{swap0to1} function in WhyML}
\label{fig:retrofit}
\end{figure}

As a result, we retrofitted the translated WhyML program by removing low-level helper functions, temporary variables, and typecasts. The retrofitting work took 1 person-day. The final WhyML program consists of 320 LOC and generates a Coq program of 403 LOC. An example of a retrofitted function is shown in Figure~\ref{fig:retrofit}.
We leave the automated
retrofitting for WhyML programs as future work.

\ProjName{} successfully verifies the retrofitted version with respect to $\mathcal{L}_1$ and $\mathcal{L}_2$ using the automated SMT solver Z3 with a total time of 1.25 seconds. And $\mathcal{L}_3$ was able to be verified in about 200 lines of Coq. 

During the experiments, we also found that most industry-grade contracts that are hard to reason about are not necessarily long, but involve complex data structures and financial models. The complexity is needed to ensure either the integrity of the free and fair market against all possible trader behavior, or the correctness of crucial data structures that involve frequent re-structuring, e.g., deletion, insertion, sorting in a linked list, etc.

\para{Compiler front-end bugs.}
In Section~\ref{sec:generatingir}, we mention that \ProjName{} can
also be used to detect
bugs or any semantic changes due to versioning ~\cite{array_creation_memory_overflow_bug}
in the compiler front-end.
Here, we describe how we use
\ProjName{} to detect a bug in 
\texttt{Solc}, the Solidity compiler,
before the  $0.6.5$ version.
Consider the code snippet in Figure~\ref{fig:arr_overflow} in which function \texttt{f} creates a dynamically-sized array in memory based on a \texttt{length} parameter. A simplified Yul IR compiled by a faulty compiler is shown in Figure~\ref{fig:arr_overflow_ir}.

\begin{figure}[t]
\begin{center}
\begin{lstlisting}[style=solidity]
contract C {
    function f(uint length) public {
        uint[] memory x = new uint[](length);
        /* other operations */
    }
}
\end{lstlisting}
\end{center}
\vspace{-10pt}
\caption{A contract with a bug of array creation overflow}
\label{fig:arr_overflow}
\end{figure}

\begin{figure}[t]
\begin{center}
\begin{lstlisting}[style=solidity]
function allocate(size) -> memPtr {
    memPtr := mload(64)
    let newFreePtr := add(memPtr, size)
    // protect against overflow
    if or(gt(newFreePtr, 0xffffffffffffffff), lt(newFreePtr, memPtr)) 
       { revert(0, 0) }
    mstore(64, newFreePtr)
}

function create_memory_array(length) -> memPtr {
    size := mul(length, 0x20)
    size := add(size, 0x20)
    memPtr := allocate(size)
    mstore(memPtr, length)
}

// main function
function f(length)  {
    let vloc_x_10_mpos := create_memory_array(length)
}
\end{lstlisting}
\end{center}
\vspace{-10pt}
\caption{Yul IR (simplified) compiled by a faulty compiler}
\label{fig:arr_overflow_ir}
\end{figure}

\begin{figure}[t]
\begin{center}
\begin{lstlisting}[language=Why3]
let function create_memory_array (st_c: evmState) (length : int) : (evmState)
(* Compiler comes with overflow check for parameters; therefore length variable does not over/underflow. *)
require { 0 <= length < 0xffff_ffff_ffff_ffff }
= let st_g: ref evmState = ref st_c in
  let ret: ref int = ref 0 in
  let st_g := push (!st_g) (Cons length (Cons 0x20 Nil)) in
  let st_g := mul (!st_g) in
  let ret := pop (!st_g) in
  (* ensure the size variable is within bound after MUL *)
  assert { 0 <= (!ret) < 0xffff_ffff_ffff_ffff }
  let st_g := push (!st_g) (Cons (!ret) (Cons 0x20 Nil)) in
  let st_g := add (!st_g) in
  let ret := pop (!st_g) in
  (* ensure the size variable is within bound after ADD *)
  assert { 0 <= (!ret) < 0xffff_ffff_ffff_ffff }
  (* ... other operations ... *)
\end{lstlisting}
\end{center}
\vspace{-10pt}
\caption{The translated WhyML function of \texttt{create\_memory\_array}}
\label{fig:arr_overflow_whyml}
\end{figure}

Inspecting the compiled \texttt{create\_memory\_array} function, we can see that since it does not check the length of the dynamically allocated array, overflow could occur by invoking \texttt{f}. \ProjName{} detects this compiler bug with the \texttt{overflow} pattern annotation. The translated WhyML function (simplified) is shown in Figure~\ref{fig:arr_overflow_whyml}.
With these fine-grained specifications embedded, \ProjName{} returns counterexamples that violate the two conditions on line 10 and 15 in Figure~\ref{fig:arr_overflow_whyml}, such as $\texttt{length} = 2^{64}/32$. This indicates that although the input parameter \texttt{length} is within a maximum bound $b$, overflow can occur after the \texttt{MUL} and \texttt{ADD} instructions (line 11 and 12 in Figure~\ref{fig:arr_overflow_ir}). The overflow checks embedded by \ProjName{} thus allow us to detect the anomalies after \texttt{MUL} and \texttt{ADD}. To eliminate this bug, a potential solution is to limit the size of the \texttt{length} parameter such that its value after \texttt{MUL} and \texttt{ADD} operations are still kept in bound. This solution is implemented in \texttt{Solc} $0.6.5$~\cite{array_creation_memory_overflow_bug}. 
To reflect this fix in WhyML, we can modify the pre-condition on line 3 with the following
\begin{lstlisting}[language=Why3]
    requires { 0 <= length < 0xffff }
\end{lstlisting}
and the verification conditions pass successfully. 
By inspecting fine-grained EVM intermediary states, \ProjName{} allows compiler front-end developers to peek into the execution of EVM, enabling the development of a more robust front-end.

\begin{table}[t]
    \centering \small
    \caption{Comparison of verification tools based on techniques and capabilities.}
    \label{tab:tools}
    \begin{tabular}{|r|c|c|c|c|}
        \hline
        \textbf{Techniques} & \textbf{Securify} & \textbf{VeriSol} & \textbf{Solc-verify} & \textbf{\ProjName{}} \\
        \hline
        \hline
        Static analysis & \green & \red & \red & \green \\
        \hline
        Symbolic execution & \red & \green & \red & \red \\
        \hline
        Semantic formalization & \green & \green & \red & \green \\
        \hline
        Model checking & \red & \green & \red & \red \\
        \hline
        Mechanized proof & \red & \green & \green & \green \\
        \hline
        IVL & \red & \red & \green & \green \\
        \hline
        Machine learning & \red & \red & \red & \green \\
        \hline
        \hline 
        \textbf{Capabilities} & & & & \\ 
        \hline
        \hline
        T1 & \red & \green & \green & \green \\
        \hline
        T2 & \red & \green & \green & \green \\
        \hline
        T3 & \red & \red & \green & \green \\
        \hline
        T4 & \red & \red & \red & \green \\
        \hline 
        T5 & \green & \red & \red & \green \\
        \hline 
        T6 & \red & \red & \red & \green \\
        \hline 
        Detecting compiler bugs & \red & \red & \red & \green \\
        \hline 
    \end{tabular}
\end{table}

\para{Cross-comparison.} 
We also compare \ProjName{} with Securify~\cite{tsankov2018securify}, Solc-verify\cite{hajdu2019solc}, and VeriSol~\cite{wang2019formal}, in terms of the verification techniques and capabilities of each framework,
shown in Table \ref{tab:tools}. Securify performs static analysis on Solidity contracts to check against pre-defined patterns. Securify supports many useful patterns and we found in our experiment that it could handle more T5 properties than \ProjName{}; however, performing security patterns alone is insufficient when many crucial properties are user-defined and contract-specific. 
Due to time-limits and engineering challenges, we did not experiment with all of the tools available. However at a high level, while VeriSol can verify most of the properties in the benchmark, it does not provide an expressive way for the user to specify custom properties. As for Solc-verify, although it provides an annotation system, its expressiveness is limited and the source-level approach relies on the correctness of the \texttt{solc} compiler. 
Compared to these tools, \ProjName{} is the most versatile framework that can verify all six types of properties. Besides an expressive annotation system, \ProjName{} allows the user to use security patterns and loop invariant inference to enhance automaton, reducing the proof burden on the user side. In addition, \ProjName{}'s IR-level approach and the fine-grained EVM model ensures
that the verified guarantees still hold on EVM
and enables the detection of errors in compiler front-end. 
Finally, \ProjName{}'s use of the Why3 IVL enables porting the generated VCs to multiple SMT solvers as well as the Coq proof assistent, facilitating flexible and compositional proofs. Overall, \ProjName{} is a capable verification framework when dealing with various proof tasks from real-world contracts. 

%% file: sections/8_related.tex
\section{Related Work}
\label{sec:RelatedWork}

\para{Pattern-based static analysis over smart contracts.}
Pattern-based techniques have been applied to detect various pre-defined vulnerability patterns in smart contracts. Vandal~\cite{brent2018vandal} and Mythril~\cite{feist2019slither} provide a static analysis framework for semantic inference. They transform the original program into custom IRs, which enable
the extraction of dataflow facts and dependency relations. Datalog-based approaches~\cite{tsankov2018security}, such as Securify~\cite{tsankov2018securify}, decompile EVM bytecode and use optimizations on semantic database queries to match vulnerability patterns. MadMax~\cite{grech2018madmax} identifies gas-related vulnerabilities and provides a control-flow-based decompiler for declarative pattern programming. 
However, pattern-based techniques may not
provide any security guarantees due to false positives and cannot express complex patterns such as
financial model related properties,
which are both supported by \ProjName{}.

\para{Model checking for smart contracts.}
Model checking techniques \cite{sergey2018temporal, nehai2018model, ahrendt2019verification, mavridou2018tool} can be applied to verify access control and temporal-related properties in smart contracts. 
However, these model checking-based approaches
may not be sound
as they tend to abstract away the contract source code and low-level execution semantics. In addition, these approaches often require contracts to be written using explicit state definitions and transitions, which can be non-intuitive to developers.
On the contrary, \ProjName{}'s 
EVM model ensures that the verified guarantees hold on the EVM,
and that
the source-level annotation system  can seamlessly embed into the source code of smart contracts and does not require any change to the program's logic.

\para{Automated verification frameworks for smart contracts.}
Existing verification frameworks use SMT solvers to encode and prove the security properties of smart contracts.  Oyente~\cite{luu2016making} employs symbolic execution to check user-defined specifications of each function on its custom IR. 
VerX~\cite{permenev2019verx} provides refinement-based strategies to contract-level invariant verification. Solc-verify~\cite{hajdu2020smt} also provides an encoding of Solidity data structures for reasoning and modeling of its semantics. Solidity compiler's SMT solver~\cite{alt2018smt} enables native keywords to be used for specification and symbolic execution of functions.
Although these frameworks achieve great automation, their capabilities are constrained by SMT solvers, which cannot express and solve clauses beyond first-order logic. On the contrary, \ProjName{} combines SMT-based verification techniques with manual verification through the Coq proof assistant. This combination enables \ProjName{} to automatically prove VCs expressed in
first-order logic and port complex
and contract-specific VCs  
to the Coq proof assistant.

\para{Semantic formalization and mechanized proofs
for smart contracts.} 
There have been extensive works on the formalization of EVM semantics~\cite{hirai2017defining, grishchenko2018semantic, amani2018towards},
which can be used to reason about EVM bytecode
but have not been applied to verify real-world smart contracts.
The K Framework~\cite{hildenbrandt2018kevm} introduces an executable EVM formal semantics and has been shown
to enable the verification of smart contracts.
This work, however, restricts the  specifications 
and the reasoning to the bytecode level,
which discards certain human-readable information, such as variable names
and explicit control flows,
making it hard for developers to  write specifications
and proofs.
Li et al.~\cite{li2019towards}
provide a formalization of smart contracts in Yul and prove the correctness of a set of benchmark contracts in Isabelle; however, the methodology presented mostly relies on manual efforts to formulate specifications and construct proofs, rendering the verification approach unscalable to complex smart contracts.  
On the contrary, \ProjName{} allows users to express properties using our
annotation system at the source level while, at the same time, offering a
highly automated verification engine, as well as
a formal model of the EVM semantics,  ensuring that 
the verified guarantees still hold over the EVM.

%% file: sections/9_conclusion.tex
\section{Conclusion}
\label{sec:Conclusion}

\ProjName{} is a versatile framework 
for specifying and verifying security properties
for  real-world smart contracts.
\ProjName{} provides an annotation system
for writing specifications
at the source code level,
translates annotated smart contracts
into annotated Yul IR programs,
and then generates VCs
using an EVM model implemented in WhyML.
Generated VCs can be discharged to SMT solvers
and the Coq proof assistant.
To further reduce the proof burden,
\ProjName{} uses 
automation techniques, including loop invariant inference and
static analysis for security patterns.
We have successfully verified 158 security properties in six types
for 12 benchmark contracts and a real-world DeFi contract. 
We expect that \ProjName{}
will become a critical building block
for developing secure and trustworthy smart contracts in the future.

\section{Acknowledgments}
We would like to thank Shoucheng Zhang for his tremendous support of our work on building trustworthy blockchain systems and smart contracts. We also thank Zhong Shao, Vilhelm Sj\"oberg, Zhaozhong Ni,
and Justin Wong for their helpful comments and suggestions that improved this
paper and the implemented tools. 
This research is based on work
supported in part by NSF grants CCF-1918400,
a Columbia-IBM Center Seed Grant Award,
and an Arm Research Gift. Any opinions, findings, conclusions, or recommendations expressed herein are those of the authors
and do not necessarily reflect those of the US Government, NSF, IBM, or Arm.